# Effect of Cell Residence Time Variance on the Performance of an Advanced Paging Algorithm

I.Z. Koukoutsidis,[*] P.I. Papaioannou, and M.E. Theologou

**Abstract**

The use of advanced sequential paging algorithms has been suggested as a means to reduce the signaling cost in future mobile cellular networks. In a proposed algorithm (Koukoutsidis and Theologou, 2003), the system can use the additional information of the last interaction cell combined with a mobility model to predict the short-term location probabilities at the time of an incoming call arrival. The short-term location probabilities reduce the uncertainty in mobile user position and thus greatly improve the search. In this paper, an analytical model is derived that allows for a general distribution of cell residence times. By considering a Gamma distribution, we study the effect of the variance of cell residence times and derive useful results on the performance of the algorithm.

## 1 Introduction

In mobile communications networks, the paging procedure is used to implement the final and subtle task of locating a mobile terminal (MT). The search is conducted both in the fixed and the wireless network, and consists of sending broadcast polling messages to all cells where the MT might be located. In current systems, the search is performed in registration zones called location areas (LAs). Usually an LA consists of a few tens of cells, which leads to a great amount of redundancy in signaling messages sent to retrieve a single user. The message replication imposes a heavy burden on both the fixed and the wireless network infrastructure. In the fixed network, encumbered elements are mainly switches, location databases and base station queues. In the radio interface, redundant page messages consume critical wireless bandwidth of the dedicated broadcast channels. More efficient use of signaling could serve to multiplex other information, or grant channels for use to transfer of data and other resource demanding applications.

---

This work was performed while the authors were at the School of Electrical and Computer Engineering of the National Technical University of Athens.

[*]Corresponding author; e-mail: `i.koukoutsidis@di.uoa.gr`

A lot of research has focused on optimizing the search process and reducing the number of polling messages sent. Usually optimization is constrained on keeping the search delay under a small reasonable value, since paging is a real-time intensive operation. In a method proposed in [1], the use of an advanced sequential paging algorithm is suggested to reduce the paging cost in a wireless network. The algorithm exploits last interaction cell information and uses a known motion model to predict the short-term movements of an MT, so as to estimate its position at the time of the next call arrival. The short-term location probabilities reduce the uncertainty in mobile user position and thus greatly improve the search. Paging is performed sequentially in order of decreasing location probabilities, as this is the optimal strategy that minimizes average costs [2]. As discussed in [1], optimal algorithms can be used to group cells into paging areas (PAs), so that delay constraints are satisfied.

A simple exponential cell residence time distribution was used to study the algorithm and derive results in [1]. However, a generalized Gamma distribution can better fit the real cell residence times, as illustrated in [3]. Besides better modeling the real system, a Gamma distribution also permits us to study the effects of variance of the cell dwell times in the algorithms performance. A usual paging algorithm that queries cells in order of decreasing location probabilities cannot be studied analytically to view the effects of variance (except via simulation), as only the mean of the distribution is considered. However, an analytical formulation that allows generally distributed cell residence times can be derived to model the proposed scheme in [1] and study the variance. The absolute minimum paging cost is used as performance metric (i.e. delay constraints are not considered), without significant loss of generality. In addition, a single network area is considered here, although a more complicated analysis would permit to study a system partitioned into location areas, as in current systems.

## 2 Analytical model

A mathematical location prediction model is expressed by a time-varying conditional distribution $p(i|u,t)$, which gives the probability of a user being located in cell $i$ at time $t$, provided he was in cell $u$ at $t=0$. The setting of the timer at $t=0$ refers to a last interaction that occurred in cell $u$. A last interaction (LI) is defined as any transaction that can provide incidental information regarding the cell position of a mobile user, such as incoming or outgoing call setup, location update or registration at switch-on, execution of messaging services, etc. We also define a paging interval as the period between the last interaction and the next incoming call arrival. If the paging interval is not too long, the conditional probabilities are much more concentrated in a small portion of the system area and thus we can largely diminish the uncertainty in mobile user position. As a result, the gains attained by applying an optimal or sub-optimal paging algorithm are vastly improved.

Assume the service area consists of $n$ cells and that cell residence times of an MT are independent and identically distributed random variables with mean $1/\lambda_m$ and distribution function denoted by $F_m(t)$. By considering the stepwise transitions of the MT, we can construct an embedded Markov chain to describe user movements in the area. Each state of the diagram corresponds to a single cell. The transition matrix of the detailed Markov walk is then an $n \times n$ matrix P= $(p_{ij})$. An element $p_{u,i}^k$ of the $k$-th powered matrix $P^k$ gives the probability that an MT initially in cell $u$ will move to cell $i$ after $k$ steps.

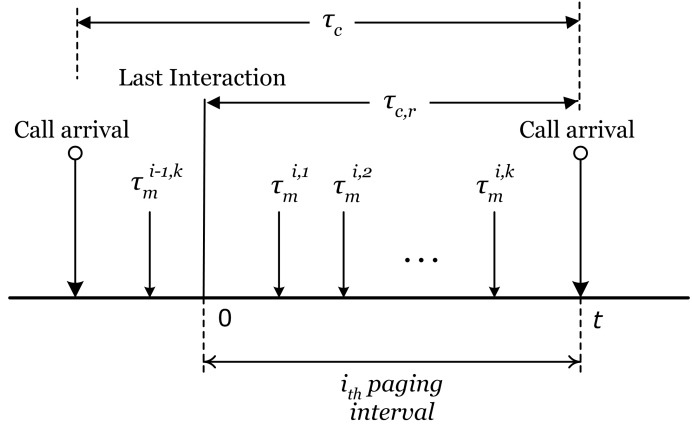

Figure 1: Timing diagram of a paging interval

In order to compute the probability $p(i|u,t)$, we need to derive the exact number of cells a user has crossed by time $t$. Let us study the timing diagram of Fig. 1 that describes a sequence of events in the $i$-th paging interval. The user may cross a number of $k$ cells between the last interaction and the next incoming call arrival. The cell boundary crossing instants in an interval $i$ are denoted by $\tau_m^{i,1}, \tau_m^{i,2}, \ldots, \tau_m^{i,k}$. Since cell dwell times are independent and identically distributed, the cell crossing instants form a *delayed renewal process* $\{\tau_m^{i,n}, n \geq 0\}$. The initial delay is the residual time in a cell measured from the last interaction after the previous boundary crossing $\tau_m^{i-1,k}$. According to renewal theory [4], this has the distribution

$$\hat{F}_m(t) = P(\tau_m^{i,1} \leq t) = \lambda_m \int_0^t (1 - F_m(x))dx \ . \tag{1}$$

In general, the distribution of the renewal points is given by the convolution

$$P(\tau_m^{i,\ell} \leq t) = \hat{F}_m(t) * F_m^{(\ell-1)}(t), \quad \ell \geq 1 \ , \tag{2}$$

where $F_m^{(\ell)}(t)$ denotes the $\ell$-fold convolution of $F_m(t)$ with itself. Now if $N_c(t)$ denotes the number of cells crossed by the MT in a period $t$, we may observe that $P(N_c(t) \geq k) = P(\tau_m^{i,k} \leq t)$. From this it is easily obtained that

$$P(N_c(t) = k) = \begin{cases} 1 - \hat{F}_m(t), & \text{if } k = 0 \\ \hat{F}_m(t) * [F_m^{(k-1)}(t) - F_m^{(k)}(t)], & \text{if } k \geq 1 \end{cases}. \quad (3)$$

If the call arrivals to each mobile terminal form a Poisson process with rate $\lambda_c$, the incoming call interarrival time is an exponentially distributed random variable with density $f_c(t) = \lambda_c e^{-\lambda_c t}$. Due to the memoryless property of the exponential distribution, the remaining time $t_{c,r}$ from the last interaction to the next call arrival is also exponential with the same mean.

Using the continuous version of the total probability theorem, the probability of crossing $k$ cells in a paging interval is

$$a(k) = \int_0^\infty P(N_c(t) = k) f_c(t) dt, \quad (4)$$

which after some manipulations yields

$$a(k) = \begin{cases} 1 - \dfrac{1 - f_m^*(\lambda_c)}{p}, & \text{if } k = 0 \\ \dfrac{1}{p}[1 - f_m^*(\lambda_c)]^2 [f_m^*(\lambda_c)]^{k-1}, & \text{if } k > 0 \end{cases}, \quad (5)$$

where $p$ is the *call-to-mobility ratio* $\lambda_c/\lambda_m$ and $f_m^*(\lambda_c)$ is the Laplace transform of the dwell time density evaluated at $\lambda_c$. This is a known formula derived using a different approach initially in [5], and is used in other contexts as well.

From there it is easy to deduce that the probability of the user residing in cell $i$ when the next call arrives is

$$p(i|u) = \sum_{k=0}^\infty a(k) p_{u,i}^k. \quad (6)$$

We use a two-parameter Gamma distribution to model cell dwell times. Its Laplace transform is given as

$$f_m^*(s) = \left(\frac{\gamma \lambda_m}{s + \gamma \lambda_m}\right)^\gamma, \quad (7)$$

where $\gamma$ is the *shape* parameter. The variance of the distribution is given as $V = 1/\gamma \lambda_m^2$. Therefore for the same mean, a large value of $\gamma$ gives a smaller variance, and vice-versa. The form of the distribution for different variances is illustrated in Fig. 2. When $\gamma = 1$, it results in an exponential distribution.

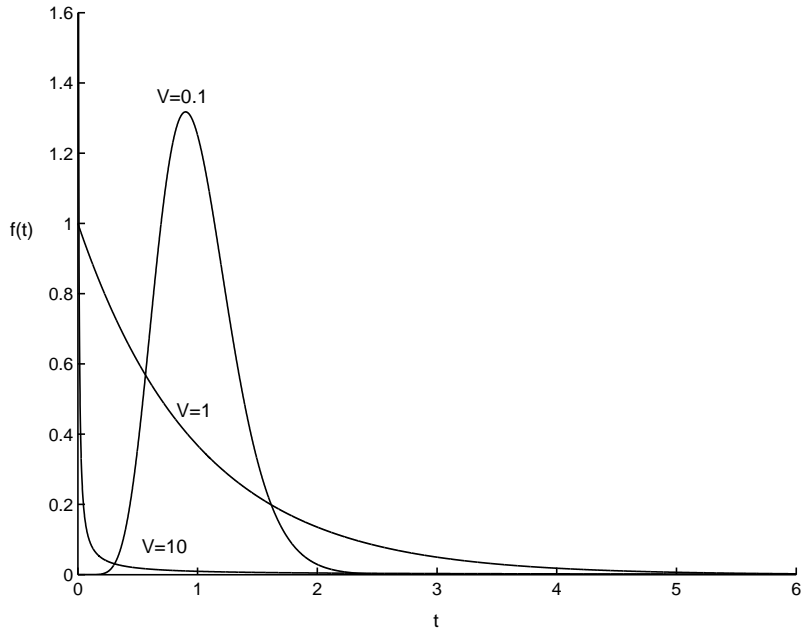

Figure 2: Density function of the Gamma distribution for different variances and mean value 1

In order to calculate the cost of an optimal sequential strategy, we sort probabilities $p(i|u)$ in decreasing order, say $q_j$, $j = 1, \ldots, n$, such that $q_1 \geq q_2 \geq \cdots \geq q_n$. Naturally, it holds that $\sum_j q_j = 1$. Then the mean paging cost is derived, depending on the LI cell $u$, as

$$C_P(u) = \sum_{j=1}^{n} j q_j \ . \tag{8}$$

## 3 Numerical results

We investigate numerical results for a hexagonal service area consisting of 31 cells. A user is allowed to move anywhere in the area according to a symmetric random walk model, i.e. upon leaving one cell, he visits each of the neighboring cells with the same probability. Both the service area and the mobility model are indicatively chosen and do no affect the nature of the results.

The paging cost varies when different last interaction cells are considered (since their placement in the service area is different), leading to a lower cost for near-border cells. In the paper, the total cost for the whole area is derived by taking into consideration the steady-state probability of the user residing in each cell. These are computed by solving the system $\pi \mathrm{P} = \mathrm{P}$, and are also the same for the continuous model since the mean residence times are equal [4]. The total cost is calculated by taking each cell as the LI cell, as

$$C_T = \sum_{u=1}^{n} \pi(u) C_P(u) ,  \quad (9)$$

where $\pi(u)$ is the steady-state probability of the user residing in cell $u$.

An important remark that is not very apparent from the analytical model is that the paging cost is independent of the individual values of the call and mobility rate. Only the ratio of the two values affects the cost function and is considered here. Results are illustrated in Fig. 3 and describe the (mean) paging cost as a function of the call-to-mobility ratio (CMR), for different variances of the cell residence time distribution.

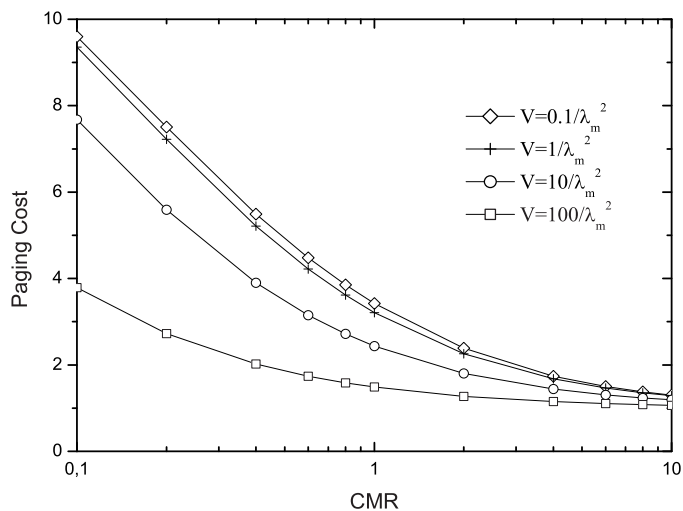

Figure 3: Paging costs by the advanced paging algorithm as a function of the CMR, for different variances of the cell residence time distribution

It is anticipated that as the CMR increases, the algorithm depicts a better performance. This happens because the relative mobility of the terminal is reduced and the location database of the system is updated more often about user positions. As a result, the conditional location probabilities are higher near the last interaction cell, and the paging cost is decreased. In the worst case, when the paging interval is infinite, conditional probabilities become equal to the steady-state probabilities and savings - compared to the classical flooding strategy - are minimized.

The variance affects the cost function in the following manner. When the variance is high, the paging cost is lower, i.e. the algorithm performs better. A significant decrease can be attained for very high variances —which can be greater than 50%, as shown in the results. Fig. 2 implies that a high variance leads to smaller cell dwell times with increased probability, but also very high values may be observed (even with very small probability). Hence it seems that large dwell times weigh more on the paging cost, even though shorter ones occur more often. On the other hand, slightly worse results are observed for smaller variances, compared to the exponential case.

The effect of large variance is more significant for a CMR lower than one; intuitively, as the mobile moves faster, the effect of having larger residence times plays a greater role in the overall performance. When the CMR is higher, the scheme has an improved performance by itself, which undermines the impact of the variance. For very high CMR values, the impact of the variance is negligible.

# 4 Ending notes

We have examined the effect of cell residence time variance on the performance of an advanced sequential paging algorithm. The algorithm itself is able to significantly decrease the access cost compared to traditional strategies, by exploiting both the short-term location probabilities and a sequential search mechanism. It has been demonstrated that the performance of the scheme is significantly affected by the actual residence time distribution; more specifically, the access cost can be greatly decreased in case of large variance of the distribution, especially when the call-to-mobility ratio is smaller than one.

Different versions of an advanced paging may exist. For example, a delay constraint can be set; after a certain delay, the sequential algorithm would be trasformed to a flooding scheme. The case of a delay constraint has not been examined in connection with the variance. However, the results change only in scale (i.e. smaller delay $\rightarrow$ higher cost), and the impact of the variance is the same. In addition, the algorithm can be applied to a system partitioned into location areas, as taken into consideration in the analysis in [1]. In this case, by applying the condition that a user does not abandon a location area before the next call arrival (since, if abandoned, a location update is performed inevitably), the location probabilities simply scale to higher values. Hence in conclusion, the main conclusions from this work apply to different versions of an advanced paging algorithm.

# References

[1] I.Z. Koukoutsidis, M.E. Theologou,"Last interaction based paging in mobile cellular networks", in Proc. Personal Wireless Communications Conference (PWC 2003), Venice, Italy, September 2003.

[2] C. Rose, R. Yates, "Minimizing the average cost of paging under delay constraints", *Wireless Networks*, vol. 1, no. 2, pp. 211–219, 1995.

[3] M.M. Zonoozi and P. Dassanayake, "User mobility modeling and characterization of mobility patterns", IEEE Journal on Selected Areas in Communications, vol. 15, no. 7, pp. 1239–1252, September 1997.

[4] S.M. Ross, *Stochastic Processes*, 2nd edn., John Wiley & Sons, New York, 1996.

[5] Y.-B. Lin, "Reducing location update cost in a PCS network", IEEE/ACM Transactions on Networking, vol. 5, no. 1, pp. 65-73, February 1997.